\documentclass[12pt,preprint]{aastex}

\slugcomment{Accepted by ApJ in May 2013}

\shorttitle{Photometric Redshifts for QSO in multi band surveys}
\shortauthors{Brescia et al. 2013}

\begin{document}


\title{Photometric redshifts for Quasars in multi band Surveys}


\author{M. Brescia\altaffilmark{1,2}}
\affil{INAF-Astronomical Observatory of Capodimonte, via Moiariello 16, I-80131 Napoli, Italy}
\email{brescia@oacn.inaf.it}

\author{S. Cavuoti\altaffilmark{2}}
\affil{Department of Physics, University Federico II, via Cinthia 6, I-80126 Napoli, Italy}

\author{R. D'Abrusco\altaffilmark{3}}
\affil{Harvard Smithsonian Center for Astrophysics, Cambridge, MA, USA}

\author{G. Longo\altaffilmark{2,4}}
\affil{Department of Physics, University Federico II, via Cinthia 6, I-80126 Napoli, Italy}

\author{A. Mercurio\altaffilmark{1}}
\affil{INAF-Astronomical Observatory of Capodimonte, via Moiariello 16, I-80131 Napoli, Italy}


\altaffiltext{1}{INAF-Astronomical Observatory of Capodimonte, via Moiariello 16, I-80131 Napoli, Italy}
\altaffiltext{2}{Department of Physics, University Federico II, via Cinthia 6, I-80126 Napoli, Italy}
\altaffiltext{3}{Harvard Smithsonian Center for Astrophysics, Cambridge, MA, USA}
\altaffiltext{4}{Visiting Associate, California Institute of Technology, Pasadena, CA, USA}


\begin{abstract}
MLPQNA stands for Multi Layer Perceptron with Quasi Newton Algorithm and it is a machine learning method which can be used to cope with regression and classification problems on complex and massive data sets. In this paper we give the formal description of the method and present the results of its application to the evaluation of photometric redshifts for quasars.
The data set used for the experiment was obtained by merging four different surveys (SDSS, GALEX, UKIDSS and WISE), thus covering a wide range of wavelengths from the UV to the mid-infrared. The method is able i) to achieve a very high accuracy; ii) to drastically reduce the number of outliers and catastrophic objects; iii) to discriminate among parameters (or features) on the basis of their significance, so that the number of features used for training and analysis can be optimized in order to reduce both the computational demands and the effects of degeneracy. The best experiment, which makes use of a selected combination of parameters drawn from the four surveys, leads, in terms of $\Delta z_{norm}$ (i.e. $(z_{spec}-z_{phot}) / (1+z_{spec})$), to an average of $\Delta z_{norm} = 0.004$, a standard deviation $\sigma = 0.069$ and a Median Absolute Deviation $MAD = 0.02$ over the whole redshift range (i.e. $z_{spec} \leq 3.6$), defined by the 4-survey cross-matched spectroscopic sample. The fraction of catastrophic outliers, i.e. of objects with photo-z deviating more than $2\sigma$ from the spectroscopic value is $< 3\%$, leading to a $\sigma = 0.035$ after their removal, over the same redshift range. The method is made available to the community through the DAMEWARE web application.
\end{abstract}





\section{Introduction}
\label{qso:intro}

Photometric redshifts (hereinafter photo-z) provide  an estimate of the redshift of sources obtained using photometry instead of spectroscopy. They are in fact driven by: (i) the shape of the broadband continuum of the object's spectroscopic emission, and (ii) by a limited number of strong spectral features (i.e. the one at 4000 \AA,  the Ly$\alpha$ forest and the Lyman limit), which are still recognizable after the integration of the Spectral Energy Distribution (SED) sampled by the filter's transmission function.

At the price of lower accuracy, photo-z offer several advantages with respect to their spectroscopic counterparts: (i) being derived from intermediate/broad band imaging, photo-z are much more effective in terms of observing time; (ii) they may allow to probe objects much fainter than the spectroscopic flux limit and (iii) under specific conditions, they allow to correct some biases, such as those encountered at high redshift where, as it has been noticed \citep{fernandez2001}, spectroscopy is pushed to its limits both by the low signal-to-noise ratio (SNR) in the spectra and by the fact that, in many cases, even when a good signal-to-noise ratio is achieved, the lack of features in the observed spectral range may undermine the estimation of a trustworthy redshift \citep{lanzetta1998}.

The latter aspect becomes crucial when photometric redshifts methods are applied to quasars (QSO) and, in particular to the construction and characterization of the large, complete samples which are required by modern cosmology. In fact, quasar samples have always been, and still are, constructed either by compiling lists of  more or less serendipitous discoveries  obtained with different techniques and selection criteria \citep{veron2000}, or via a two-step process where the first one consists in the identification of QSO candidates from multi-wavelength surveys, and the second requires the spectroscopic validation of the candidates. In practice, due to the large amount of observing time required by spectroscopy, the latter step is usually optimized by applying the spectroscopic validation procedure just to a more or less significant subsample of the candidates, and then by extrapolating the resulting statistics to the whole sample. Modern surveys are usually so deep and extensive that the number of candidates rapidly becomes too large to be handled with the latter approach. On the other hand, modern multi-wavelength digital surveys also provide such a wealth of information (multi-band high accuracy photometry) that it becomes feasible to approximate the SED of objects over a quite large range of redshifts \citep{richards2001a,richards2001b,budavari2001,wolf2004}, thus minimizing the need for spectroscopic follow-up.

In the last few years it has in fact been demonstrated that, after having provided an accurate enough photometry and significant wavelength coverage, it is possible to obtain samples of photometrically selected quasars matching the low contamination and high completeness \citep{dabrusco2009, bovy2012} required by many fields of modern cosmology. The relevance of these photometric samples will increase more and more in the near future, when the new generation of deeper and more accurate surveys will allow to access larger and more complete samples of QSOs. These \textit{photometric} samples are in fact already being used for a variety of applications such as the measurement of the integrated Sachs--Wolfe effect \citep{giannantonio2008}, the cosmic magnification bias \citep{scranton2005},  the clustering of quasars on large \citep{myers2006} and small \citep{hennawi2006} scales, to quote just a few. Since both candidate selection and photometry redshift estimates are performed on the same data (colors in many bands), it is also apparent that for the same samples, photometric data alone should carry enough information to characterize in an almost univocal way the SED and therefore also to derive accurate estimates of photometric redshifts \citep{dabrusco2009,laurino2011,bovy2012}.

It goes without saying that the utility of the photometric samples goes hand in hand with the development of photo-z methods capable to provide accurate enough estimates of the redshifts.

In this paper we use a new empirical method, named Multi Layer Perceptron with Quasi Newton
Algorithm  or MLPQNA, and apply it to the evaluation of photometric redshift of quasars. In section \ref{qso:sec:thedata} we discuss the datasets used for the experiments and in section  \ref{qso:sec:method} we present both a detailed description of the MLPQNA method and the statistical indicators used throughout the paper. We wish to stress that the lack of a common agreement on such indicators is among the main obstacles in comparing the performances of different methods. In section \ref{qso:sec:experiments} we describe the experiments performed in order to select the best combination of input parameters, bands and network topology. The results of these experiments are summarized and discussed in section  \ref{qso:sec:discussion}, where we also present the final performances of the best experiments. Finally, we compare our results with those available in literature and draw some general conclusions.

A short appendix provides the reader with the math behind the Quasi Newton Algorithm.

\section{The Dataset}
\label{qso:sec:thedata}

The sample of quasars, used in the experiments described in this paper, is based on the spectroscopically selected quasars from the SDSS-DR7 database (table \textit{Star} of the SDSS database). According to the spectroscopic classification index ({\it index SP} or {\it specClass}) provided in the SDSS-DR7 release \citep{schneider2010}, we selected quasars, for which a reliable measure of the spectroscopic redshifts (with {\it zConf} $>$ 0.90) is available.

We then cross-matched the SDSS quasars sample identified as point sources with clean measured photometry in all filters (\textit{ugriz}), with the latest versions of the datasets from: GALEX~\citep{martin2005}, UKIDSS~\citep{lawrence2007} and WISE~\citep{wright2010}. These three surveys observed large fractions of the sky in the ultraviolet, near infrared and middle infrared spectral intervals, respectively. After the cross matching we obtained a series of multi-band catalogues, defined as it follows.

{\bf SDSS - (DR7)}~\citep{aihara2011} has observed $\sim\!1.4\times10^4$  deg$^2$ of the sky in 5 bands (\textit{ugriz}) covering the [3551, 8931] \AA \phantom{ }wavelengths range. Photometric SDSS observations reach the limiting magnitude of 22.2 in the \textit{r} band ($95\%$ completeness for point sources; \citealt{abazajian2009}).

{\bf GALEX - (DR6/7)}~\citep{martin2005} is a 2-band survey (\textit{nuv, fuv} for near and far ultraviolet respectively) covering the [1300,3000] \AA \phantom{ }wavelength interval. The GALEX photometric survey has observed the whole sky to the near ultraviolet limiting magnitude $\textit{nuv}\!=\!20.5$.

{\bf UKIDSS - (DR9)}~\citep{lawrence2007} has been designed to be the SDSS infrared counterpart and covers $\sim$7000 deg$^2$ of the sky in the \textit{YJHK} near-infrared bands covering the $\sim$ 0.9 to 2.4 $\mu$m spectral range down to the limiting magnitude \textit{K}=18.3. The Large Area Survey (LAS) has imaged $\sim4000$ deg$^2$ (overlapping with the SDSS), with the additional \textit{Y} band down to the limiting magnitude of 20.5.

The {\bf WISE} mission~\citep{wright2010} has observed the entire sky in the mid-infrared spectral interval at 3.4, 4.6, 12, and 22 $\mu$m with an angular resolution of 6.1$^{\prime\prime}$, 6.4$^{\prime\prime}$, 6.5$^{\prime\prime}$ and 12.0$^{\prime\prime}$ in the four bands, achieving 5$\sigma$ point source sensitivities of 0.08, 0.11, 1 and 6 mJy in unconfused regions on the ecliptic, respectively. The astrometric accuracy of WISE is $\sim 0.50^{\prime\prime}, 0.26^{\prime\prime}, 0.26^{\prime\prime}$, and 1.4$^{\prime\prime}$ for the four WISE bands, respectively.

\begin{figure}
\centering
  \includegraphics[width=10.cm]{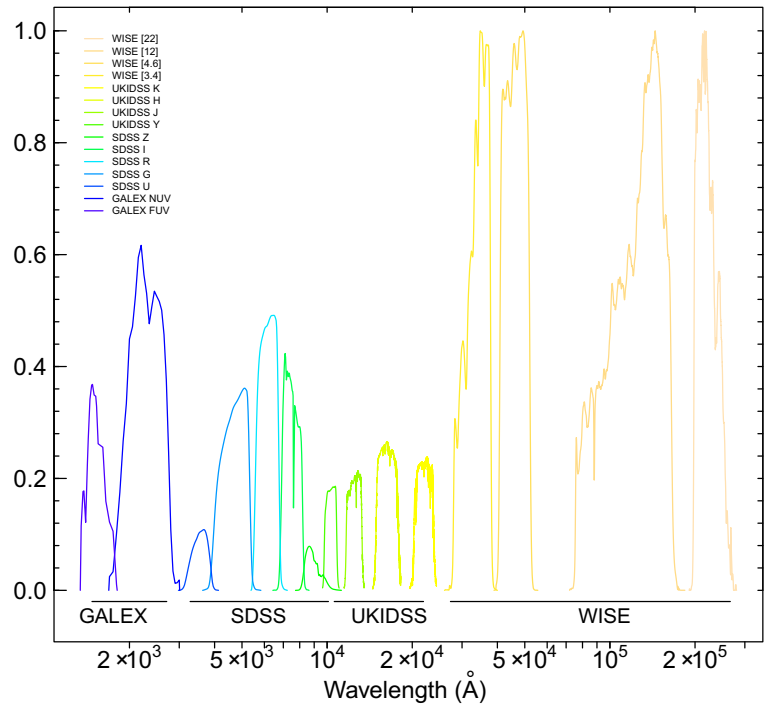}\\
  \caption{Transmission curves for all filters in the four surveys considered.\label{qso:fig:transmission_curves}}
\end{figure}

The transmission curves of all filters related with the four surveys are shown in Fig.~\ref{qso:fig:transmission_curves}. All these surveys present a large common overlap region and  overall good astrometry with comparable astrometric accuracy. In order to cross-match the catalogues we used a maximum radius $r=1.5^{\prime\prime}$ to associate the optical quasars to counterparts in each of the three catalogs.
Afterwards we rejected all sources containing one or more missing data in any of their photometric parameters. In this case with the term \textit{missing data} we mean undefined numerical values underlying either not detected or contaminated magnitude measurements. This last step is crucial in empirical methods since the presence of missing data might affect their generalization capabilities \citep{marlin2008}.

The resulting number of objects in the datasets used for the experiments are:
\begin{itemize}
\item SDSS: $\sim1.1\times10^5$;
\item SDSS $\cap$ GALEX: $\sim4.5\times10^4$;
\item SDSS $\cap$ UKIDSS:  $\sim3.1\times10^4$;
\item SDSS $\cap$ GALEX $\cap$ UKIDSS:  $\sim1.5\times10^4$;
\item SDSS $\cap$ GALEX $\cap$ UKIDSS $\cap$ WISE:  $\sim1.4\times10^4$;
\end{itemize}
An additional dataset was produced by decimating the final \textit{four-surveys} cross-matched catalogue. This dataset was used to perform the preliminary feature-selection or \textit{pruning} phase (see Sec.~\ref{qso:featureselection}) and consisted of $\sim 3.8 \times 10^3$ objects, each observed in 15 bands (4 UKIDSS, 2 GALEX, 5 SDSS and 4 WISE) and with accurate spectroscopic redshift estimates. The decimation was needed to reduce the computational time needed to perform the large number of experiments described in what follows. For some bands there were multiple measurements (i.e. magnitude measured accordingly to different definitions) and therefore we are left with a total of 43 different features.

\begin{figure}
\centering
 \includegraphics[width=12.cm]{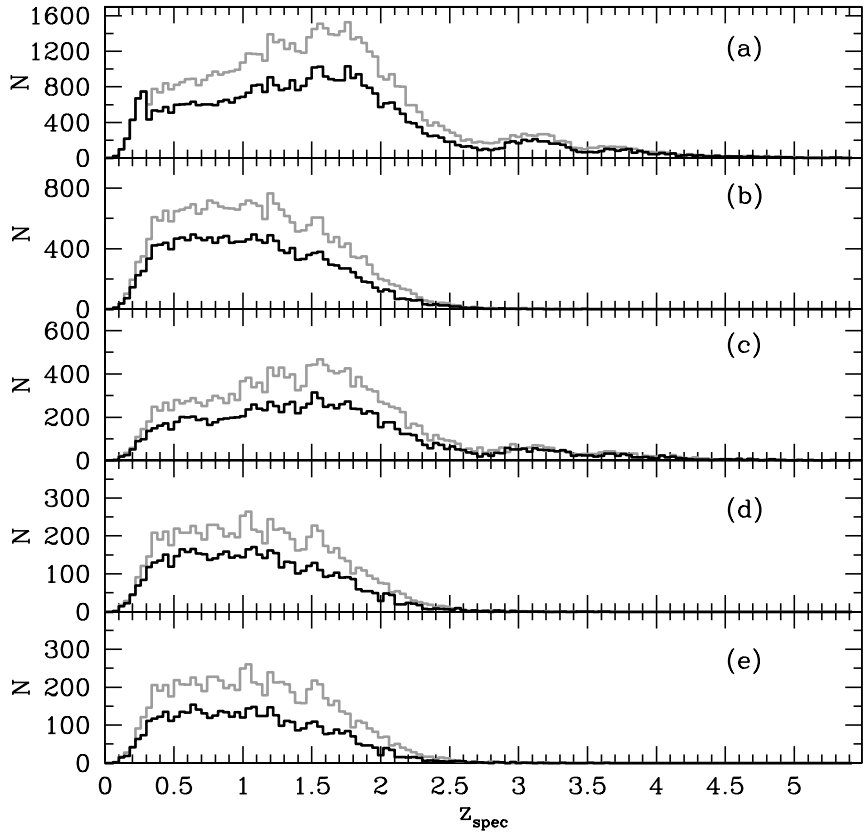}
\caption{Histograms of spectroscopic redshift distribution in the five survey cross-matched samples as derived from the SDSS spectroscopic data. (a) SDSS; (b) SDSS+GALEX; (c) SDSS+UKIDSS; (d) SDSS+GALEX+UKIDSS; (e) SDSS+GALEX+UKIDSS+WISE. Gray dotted line is the training sample. Black line is the test sample.\label{qso:fig:zspec_histograms}}
\end{figure}

Finally, in producing training and test sets we made sure that they had compatible spectroscopic redshifts distributions (see Fig.~\ref{qso:fig:zspec_histograms}).

\section{The Method}\label{qso:sec:method}
This section is dedicated to the description of the machine learning method used for the experiments. All mathematical details are reported in the Appendix.

\subsection{Multi Layer Perceptron (MLP)}
From a technical point of view, the MLPQNA method, is a Multi Layer Perceptron (MLP; \citealt{bishop2006}) neural network trained by a learning rule based on the Quasi Newton Algorithm (QNA); in other words and as it is synthesized in the acronym, MLPQNA differs from more traditional MLP's implementations in the way the optimal solution of the regression problem is found.
In previous papers, most of the characteristics of the method have been described in the contexts of both classification \citep{brescia2012a} and regression \citep{cavuoti2012b}.

According to \cite{bishop2006}, feed forward neural networks (in their various implementations) provide a general framework for representing non linear functional mappings between a set of input variables (also called features) and a set of output variables (the targets). The training of a neural network can be in fact seen as the search for the function which minimizes the errors of the predicted values with respect to the true values available for a small but significant subsample of objects in the same parameter space. This subset is also called \textit{training set} or \textit{Knowledge Base} (KB).  The final performances of a specific Neural Network (NN) depend on many factors, such as topology, the way the minimum of the error function is searched and found, the way errors are computed, as well as the intrinsic quality of training data.

The formal description of a feed-forward neural network with two computational layers is given in the Eq. \ref{qso:eq0}:

\begin{equation} \label{qso:eq0}
y_k = \sum_{j=0}^{M} w^{(2)}_{kj} g \left( \sum_{i=0}^{d} w^{(1)}_{ji} x_i \right)
\end{equation}

\begin{figure}
\centering
\includegraphics[width=6.cm]{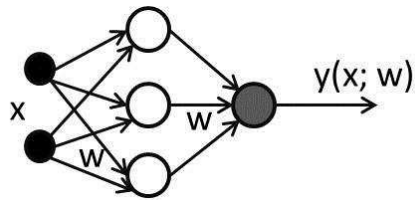}
\caption{Scheme of a Multi Layer Perceptron general architecture for two input variables, one hidden layer with three neurons and one output value.\label{qso:mlp}}
\end{figure}

Equation \ref{qso:eq0} can be better understood by using a graph like the one shown in Fig. \ref{qso:mlp}.
The input layer $\left( x_i \right)$ is made of a number of neurons (also known as perceptrons) equal to the number of input variables $\left( d \right)$; the output layer, on the other hand, will have as many neurons as the output variables $\left( k \right)$.

In the general case, the network may have an arbitrary number of hidden layers (in the depicted case there is just one hidden layer with three neurons), each of one can be formed by an arbitrary number of neurons $(M)$. In a fully connected feed-forward network each node of a layer is connected to all the nodes in the adjacent layers.
Each connection is represented by an adaptive weight $\left( w^{l}_{kj} \right)$  which can be regarded as the strength of the synaptic connection between neurons $k$ and $j$, while the response of each perceptron to the inputs is represented by a non-linear function $g$, referred to as the \textit{activation function}. All the above characteristics, the topology of the network and the weight matrix of its connections, define a specific implementation and are usually called \textit{model}.

The model, however, is only part of the story. In fact, in order to find the model that best fits the data in a specific problem, one has to provide the network with a set of examples, \textit{id est} of objects for which the final output is known by independent means. These data, already defined as \textit{training set} or \textit{Knowledge Base}, are used by the network to find the optimal model.

In our implementation we choose as learning rule the QNA, which differs from the Newton Algorithm in how the Hessian of the error function is computed. Newtonian models are variable metric methods used to find local maxima and minima of functions \citep{davidon1968} and, in the case of MLPs, they can be used to find the stationary (i.e. the zero gradient) point of the learning function. The complete mathematical description of the MLP with QNA model is reported in the appendix \ref{qso:appendixA}.

The model has been made available to the community through the DAta Mining \& Exploration Web Application REsource (DAMEWARE\footnote{\emph{http://dame.dsf.unina.it/beta\_info.html}}; \citealt{cavuoti2012c}).

\subsection{The implementation of MLPQNA}\label{qso:sec:mlpqna}
In this work we use our implementation of the QNA based on the limited-memory BFGS (L-BFGS; \citealt{byrd1994}), where BFGS is the acronym composed of the names of the four inventors \citep{broyden1970, fletcher1970, goldfarb1970, shanno1970}.

Summarising, the algorithm for MLP with QNA is the following. Let us consider a generic MLP with $w^{(t)}$ being the weight vector at time $(t)$.

\begin{enumerate}
\item Initialize all weights $w^{(0)}$ with small random values (typically normalized in $[-1, 1]$), set the constant error tolerance $\varepsilon$ and $t = 0$;
\item present to the network all training set and calculate $E(w^{(t)})$ as the error function for the current weight configuration;
\item if $t = 0$ then $d^{(t)} = -\nabla E^{(t)}$
\item else $d^{(t)}=-\nabla E^{(t-1)}+Ap+B\nu$, where $p = w^{(t+1)} - w^{(t)}$ and $\nu = g^{(t+1)} - g^{(t)}$;
\item calculate $w^{(t+1)} = w^{(t)} - \alpha d^{(t)}$, where $\alpha$ is obtained by line search equation (see Eq. \ref{qso:eq6} in the Appendix);
\item calculate $A$ and $B$ for the next iteration, as reported in Eq. \ref{qso:eq19};
\item if $E(w^{(t+1)}) > \varepsilon$ then $t = t+1$ and goto (ii), else STOP.
\end{enumerate}

As it is known, all \textit{line search} methods, being based on techniques which search for the minimum error by exploring the error function surface, are likely to get stuck in a local minimum and many solutions to this problem have been proposed \citep{floudas2005}. In order to optimize the convergence of the Gradient Descent Analysis (GDA, see Appendix), Newton's method uses the information on the second-order derivatives. By having information on the second derivatives, QNA is more effective in avoiding local minima of the error function and more accurate in the error function trend follow-up, thus revealing a \textit{natural} capability to find the absolute minimum error of the optimization problem \citep{shanno1990}.

In the L-BFGS version of the algorithm, in the case of high dimensionality (i.e. input data with many parameters), the amount of memory required to store the Hessian is too large, along with the machine time required to process it. Therefore, instead of using a complete number of gradient values to generate the Hessian, we can use a smaller number of values to approximate it.

By the way, if the convergence slows down, performances may even increase. A statement which only a first sight might seem paradoxical but, while the convergence is measured by the number of iterations, the performances depend on the number of processor's time units spent to calculate the result.

Related to the computational cost there is also the strategy adopted in terms of stopping criteria for the method. As it is known, the process of adjusting the weights based on the gradients is repeated until a minimum is reached. In practice, one has to decide the stopping condition of the algorithm. Among the possible criteria, the algorithm could be terminated when: (i) the Hessian approximation error becomes sufficiently small (by definition the gradient will be zero at a minimum); (ii) the maximum number of iterations is reached, in terms of a fixed threshold; (iii) based on the cross validation.

The cross validation can be used to monitor generalization performance during training and to terminate the algorithm when there is no more improvement. Statistically significant results come out by trying multiple independent data partitions and then averaging the performances. There are several variants of cross validation methods \citep{sylvain2010}. We, in particular, have chosen the k-fold cross validation, particularly suited in presence of a scarcity of known data samples \citep{geisser1975}. The mechanism, also known as \textit{leave-one-out},  is quite simple, since it consists in dividing the training set of $N$ samples into $k$ subsets ($k > 1$). The model is then trained on $k-1$ subsets and validated by testing it on the left out subset. This procedure is then iterated leaving out each time a different subset for validation and its mean squared error is averaged on all cycles.

For what the MLP topology is concerned, a significant contribution came from the seminal paper by Bengio \& LeCun \citep{bengio2007}. They in fact re-analysed the implications of the Haykin pseudo-theorem \citep{haykin1998}, proving that complex problems, in which the mapping function is highly non linear and the local density of data in the parameter space is very variable, are better matched by \textit{deep} networks with more than one hidden computational layer.

\subsection{Statistical Indicators}\label{qso:sec:stat}
In order to evaluate and reciprocally compare the experiments described in the next section we adopted the following definitions:\\
\begin{equation}
\mbox{bias(x)} = \frac{{\sum\limits_{i = 1}^N {x_i } }}{N}
\label{qso:eq21}
\end{equation}

\begin{equation}
\sigma(x)  = \sqrt {\frac{{\sum\limits_{i = 1}^N {\left[ {x_i  - \left( {\frac{{\sum\limits_{i = 1}^N {x_i } }}{N}} \right)} \right]^2 } }}{N}}
\label{qso:eq22}
\end{equation}

\begin{equation}
\mbox{MAD(x)} = Median\left( \left| x \right| \right )
\label{qso:eq21bis}
\end{equation}

\begin{equation}
\mbox{NMAD(x)} = 1.48 \times Median\left( \left| x \right| \right )
\label{qso:eq21tris}
\end{equation}

\begin{equation}
\mbox{RMS(x)} = \sqrt {\frac{{\sum\limits_{i = 1}^N {x_i^2 } }}{N}}
\label{qso:eqrms}
\end{equation}

where $\sigma$ is the standard deviation, MAD is the Median Absolute Deviation, NMAD the normalized MAD and RMS is the Root Mean Square. The term $x$ in all above expressions may be either $\Delta z$ defined as:\\
\begin{equation}
\Delta z = (z_{spec}-z_{phot})
\label{qso:eq20}
\end{equation}

or the normalized residuals $\Delta z_{norm}$ defined as:\\
\begin{equation}
\Delta z_{norm} = (z_{spec}-z_{phot}) / (1+z_{spec})
\label{qso:eqdznorm}
\end{equation}

\section{The experiments}
\label{qso:sec:experiments}

Our approach is based on machine learning methods and therefore, it needs to be as automatic as possible, in order to optimize the decisional support to the user (in this case the astronomer). Therefore, the results of the individual experiments as well as their comparison with others, have to be evaluated in a consistent and objective manner through an homogeneous set of statistical indicators.

In what follows we shall discuss the general experiment workflow and the outcome of the experiment phases.

\subsection{The knowledge base and model setup}\label{qso:sec:kbsetup}

For machine learning supervised methods it is common practice to use the available KB to obtain at least three disjoint subsets for every experiment: one (training set) for training purposes, i.e. to train the method in order to acquire the hidden correlation among the input features, which is needed to perform the regression; the second one (validation set) to check the training, in particular against loss of generalization capabilities (a phenomenon also known as overfitting); and the third one (test set) to evaluate the overall performances of the model.  As a rule of thumb in case of machine learning methods, these sets should be populated with respectively 60\%, 20\% and 20\% of the objects in the KB \citep{kearns1996}. In our case, however, we reduced the training+validation data amount (from 80\% to 60\%), driven by the past experience with the very accurate regression capabilities of the model also in case of smaller knowledge bases \citep{brescia2012a, cavuoti2012b}, obtaining implicitly the possibility to verify its prediction performance on a larger test set, as well as a faster execution of the training phase. Furthermore, in order to ensure a proper coverage of the KB in the Parameter Space (PS), the data objects were indeed divided up among the three datasets by random extraction, and usually this process is iterated several times to minimize the possible biases induced by fluctuations in the coverage of the PS, namely small differences in the redshift distribution of training and test samples used in the experiments.

The first two criteria used to decide the stopping condition of the algorithm, as mentioned at the end of Sec.~\ref{qso:sec:mlpqna}, are mainly sensitive to the choice of specific parameters and may lead to poor results if the parameters are improperly set. The cross validation does not suffer of such drawback; it can avoid overfitting the data and is able to improve the generalization performance of the model. However, if compared to the traditional training procedures, the cross validation is much more computationally expensive. Therefore, by exploiting the cross validation technique (see Sec.~\ref{qso:sec:mlpqna}), training and validation were indeed performed together using $\sim 60\%$ of the objects as a training + validation set, and the remaining $\sim 40\%$ as test set.

The automatized process of the cross-validation was done by performing ten different training runs with the following procedure: (i) splitting of the training/validation set into ten random subsets, each one composed by 10\% of the dataset; (ii) at each training run we applied the 90\% of the dataset and the excluded 10\% for validation.

As remarked in Sec.~\ref{qso:sec:mlpqna}, the k-fold cross validation is able to avoid overfitting on the training set \citep{bishop2006}, with an increase of the execution time estimable around $k-1$ times the total number of runs \citep{cavuoti2012b}.

In terms of the internal parameter setup of the MLPQNA, we need to consider the following topological parameters:\\
\begin{itemize}
\item input layer: a variable number of neurons, corresponding to the pruned number of survey parameters used in all experiments, up to a maximum number of 43 nodes (all available features);
\item neurons on the first hidden layer: a variable number of hidden neurons, depending on the number $N$ of input neurons (features in the dataset), equal to $2N+1$ as rule of thumb;
\item neurons on the second hidden layer: a variable number of hidden neurons, ranging from 0 (to ignore the second layer) to  $N-1$;
\item output layer: one neuron, returning the reconstructed photo-z value.
\end{itemize}

For the QNA learning rule, we heuristically fixed the following values as best parameters for the final experiments:\\
\begin{itemize}
\item step: $0.0001$ (one of the two stopping criteria. The algorithm stops if the approximation error step size is less than this value. A step value equal to zero means to use the parameter MaxIt as the unique stopping criterion);
\item res : $40$ (number of restarts of Hessian approximation from random positions, performed at each iteration);
\item dec : $0.1$ (regularization factor for weight decay. The term $dec*||network weights||^2$ is added to the error function, where $network weights$ is the total number of weights in the network, directly depending on the total number of neurons inside. When properly chosen, the generalization performances of the network are highly improved);
\item MaxIt: $8000$ (max number of iterations of Hessian approximation. If zero the step parameter is used as stopping criterion);
\item CV (k): $10$ (k-fold cross validation, with $k=10$);
\item Error evaluation: Mean Square Error (between target and network output).
\end{itemize}

With these parameters, we obtained the statistical results reported in Sec.~\ref{qso:complexityselection}.

\subsection{Selection of features}\label{qso:featureselection}

As it is known, supervised machine learning models are powerful methods for learning the hidden correlation between input and output features from training data.  Of course, their generalization and prediction capabilities strongly depend on: the intrinsic quality of data (signal-to-noise ratio), the level of correlation among different features; the amount of missing data present in the dataset \citep{ripley1996}. It is obvious that some, possibly many, of the $43$ parameters listed in Tab.~\ref{qso:tab:features} may not be independent and that their number needs to be reduced in order to speed up the computation (which scales badly with the number of features). This is a common problem in data mining and there is a wide literature on how to optimize the selection of features which are most relevant for a given purpose \citep{lindeberg1998, guyon2003, guyon2006, brescia2012d}. This process is usually called \textit{Feature selection} or \textit{pruning}, and consists in finding a compromise between the number of features (and therefore the computational time) and the required accuracy of the final results. In order to do so, we extracted from the main catalogue several subsets containing different groups of variables (features). Each one of these subsets was then analyzed separately in specific runs of the method (runs which in the data mining community are usually called experiments), in order to allow the comparison and evaluation. We wish to stress that our main concern was not only to disentangle which bands carry the most information but also, for a given band, which type of measurements (e.g. Point Spread Function, petrosian or isophotal magnitude) are more effective.

We performed a series of regression experiments to evaluate the performances obtained by the pruning of photometric quantities on the small dataset described in Sec.~\ref{qso:sec:thedata}. The pruning experiments consisted into several combinations of surveys and their features:\\
\begin{itemize}
\item a \textit{full} features experiment to be used as a benchmark for all the other experiments;
\item some \textit{service} experiments used to select the best combination of input features in order to eliminate redundancies in the flux measurements (i.e., petrosian magnitudes against isophotal magnitudes);
\item \textit{three-survey} experiments for all possible combinations of three (out of four) surveys;
\item \textit{two-survey} experiments with all possible combinations of two (out of four) surveys;
\item \textit{single-survey} experiments.
\end{itemize}

The output of the experiments consisted of lists of photometric redshift estimates for all objects in the KB. All pruning experiments were performed using $\sim 3000$ objects in the training set and $\sim 800 $ in the test set.
In Tab.~\ref{qso:tab:experiments}, we list the outcome of the experiments for the feature selection. Both $bias\left(\Delta z \right)$ and $\sigma \left(\Delta z \right)$ were computed using the objects in the test set alone. As it can be seen, among the various types of magnitudes available for GALEX and UKIDSS, the best combination is obtained using the isophotal magnitudes for GALEX and the calibrated magnitudes ($HallMag$) for UKIDSS.

Therefore at the end of the pruning phase the best combination of features turned out to be: the five SDSS $psfMag$, the two isophotal magnitudes of GALEX, the four $HallMag$ for UKIDSS and the four magnitudes for WISE.

\subsection{Magnitudes vs Colors}\label{qso:magnitudecolor}

Once the most significant features had been identified, we had to check which types of flux combinations were more effective, in terms of magnitudes or related colors. Experiments were performed on all five cross-matched datasets listed in section \ref{qso:sec:thedata}.

As it could be expected, the optimal combination turned out to be always the mixed one, i.e the one including colors and one reference magnitude for each of the included surveys (r for SDSS, nuv for GALEX, K for UKIDSS and W4 for WISE). From the data mining point of view this is rather surprising since the amount of information should not change by applying linear combinations between features. But from the physical point of view this can be easily understood by noticing that even though colors are derived as a subtraction of magnitudes, the content of information is quite different, since an ordering relationship is implicitly assumed, thus increasing the amount of information in the final output (gradients instead of fluxes). The additional reference magnitude instead removes the degeneracy in the luminosity class for a specific galaxy type.

\subsection{MLPQNA Network Topology}\label{qso:complexityselection}

The final check was about the hierarchical complexity of the network in terms of number of internal layers, whether \textit{shallow} or \textit{deep} according the definitions in \cite{bengio2007}, where \textit{deep} is referred to a feed-forward network with more than one hidden layer. The above quoted cross-matched datasets were therefore processed through both a three-layers (input + hidden + output) and a four-layers (input + 2 hidden layers + output) network.
In all cases the four-layers network performed significantly better, thus confirming the performance enhancement with \textit{deep} networks in case of a particularly complex non-linear regression cases, i.e. in case of a highly multi-variate distributions of the input parameter space.

The experiments with best results have been obtained using a four-layers network, trained on the mixed (colors + reference magnitudes) datasets and their statistics are reported in tables \ref{qso:tab:comparison1}, \ref{qso:tab:comparison2}, \ref{qso:tab:compoutliers} and \ref{qso:tab:catastrop}.

\section{Discussion and conclusions}\label{qso:sec:discussion}

 In 2002 we begun to explore the usage of MLP's for the evaluation of photo-z both for \textit{normal} galaxies and quasars \citep{tagliaferri2002}. Several years later, \citealt{dabrusco2007} used a combination of two MLP's to correct for the degeneracy introduced by the inhomogeneities in the knowledge base. Then \citealt{laurino2011} demonstrated that the subtleties in the mapping function could be more easily captured using the so-called \textit{WGE (Weak Gated Experts)} method, a hierarchical combination of MLP's each specialized in a specific partition of the parameter space, whose individual outputs were combined by an additional MLP.

Furthermore, \citealt{bengio2007} published a seminal paper which somehow has disproved the Haykin-pseudo theorem \citep{haykin1998}, pointing out that problems with a large amount of distribution irregularities  in the parameter space, are better treated by what they defined as \textit{deep} networks, i.e. networks with more than one computational (hidden) layer.  In this paper we exploited \citealt{bengio2007} findings, by using the supervised machine learning based method MLPQNA to evaluate photometric redshifts of quasars using multi-band data obtained from the cross-matching of the GALEX, SDSS, UKIDSS and WISE surveys.

In the tables \ref{qso:tab:comparison1}, \ref{qso:tab:comparison2} and \ref{qso:tab:compoutliers} we compare our best results to those presented by other authors \citep{ball2008,richards2009,laurino2011,bovy2012}, in terms of an homogeneous set of statistical indicators, defined in Sec.~\ref{qso:sec:stat}. Unfortunately, the whole set of indicators was not available for all bibliographical sources and in several cases we could only use a few quantities. Results are listed according to the combinations of surveys used in the experiment.

The best experiment, which makes use of a selected combination of parameters drawn from the four cross-matched surveys, leads to a $bias = 0.004$ and a Median Absolute Deviation $MAD = 0.02$. The fraction of catastrophic outliers, i.e. of objects with photo-z deviating more than $2\sigma$ from the spectroscopic value is $< 3\%$, leading to a $\sigma(\Delta z_{norm}) = 0.035$ after their removal (as reported in Tab.~\ref{qso:tab:catastrop}).
The larger the number of surveys (bands) used, the more accurate are the results. This result, which might seem evident, is not obvious at all, since the higher amount of information carried by the additional bands implies also a strong decrease in the number of objects which are contained in the training set and should therefore cause a decrease in the generalization performances of the network.

This result, together with the fact that MLPQNA performs well also with small KB's \citep{cavuoti2012b},
seems particularly interesting, since it has far reaching implications on ongoing and future surveys: a better photometric coverage is much more relevant than an increase of spectroscopic templates in the KB.

Concerning the performance evaluation in terms of photometric redshift reconstruction, all statistical results reported throughout this paper are referred to test data sets only. In fact, it is good practice to evaluate the results on data (i.e. the test set) which have never been presented to the network during the training and/or validation phases. The usage of \textit{test plus training} data might introduce an obvious positive systematic bias which could mask reality.

More in general, empirical methods, such as MLPQNA, have the advantage that the training set is made up of real sky objects. Hence they do not suffer from the uncertainty of having accurate templates. In this sense any empirical method intrinsically includes effects such as the filter band-pass and flux calibrations. In fact, as deeply discussed by \cite{collister2004}, one of the main drawbacks of these methods is the difficulty in extrapolating to regions of the input parameter space that are not well sampled by the training data. Therefore the efficiency of empirical methods degrades for objects at fainter magnitudes than those included in the training set, as this would require an extrapolation capability on data having properties, such as redshift and photometry, not included in the learned sample. In fact, another strong requirement of such methods is that the training set must be large enough to cover properly the parameter space in terms of colors, magnitudes, object types and redshift. In this case the calibrations and corresponding uncertainties are well known and only limited extrapolations beyond the observed locus in color-magnitude space are required. In conclusion, under the conditions described above about the consistency of the training set, a realistic way to measure photometric uncertainties is to compare the photometric redshifts estimation with spectroscopic measures in the test samples.

As it can be seen in the tables \ref{qso:tab:comparison1}, \ref{qso:tab:comparison2} and \ref{qso:tab:compoutliers}, in  all cases MLPQNA obtains very relevant results. Only in the SDSS+GALEX case, the non-normalized quantities (i.e. those referred to the error $\Delta z = z_{spec}-z_{phot}$) show a substantial agreement between our results and those by \citealt{laurino2011}. The better performances of MLPQNA in the normalized indicators (i.e. those referred to the error $\Delta z_{norm} = (z_{spec}-z_{phot}) / (1+z_{spec})$), is a consequence of the better performances of the MLPQNA method in terms of fraction of catastrophic outliers.

We wish to stress that both our four-layers MLPQNA and the \textit{WGE} method discussed in \citealt{laurino2011} take advantage of a substantial improvement in complexity with respect to the traditional three-layers MLP networks used in the literature, and therefore deal better with the complexity of the multi-color parameter space.
Average statistical indicators such as bias and standard deviation, however, provide only part of the information which allows to correctly evaluate the performances of a method and, for instance, they provide only very little evidence of the systematic trends which are observed as a sudden increase in the residuals spread over specific regions of the redshift space \citep{laurino2011}. In the worst cases, these regions correspond to degeneracies in the parameter space and, as it could be expected, the relevance of such degeneracies decreases for increasing number of bands.

For what the analysis of the catastrophic outliers is concerned, according to \cite{mobasher2007}, the parameter $D_{95} \equiv \Delta_{95}/\left(1+z_{phot}\right)$ enables the identification of outliers in photometric redshifts derived through SED fitting methods (usually evaluated through numerical simulations based on mock catalogues). In fact, in the hypothesis that the redshift error $\Delta z_{norm} = \left( z_{spec}-z_{phot}\right)/\left(1 + z_{spec}\right)$ is Gaussian, the catastrophic redshift error limit would be constrained by the width of the redshift probability distribution, corresponding to the $95\%$ confidence interval, i.e. with $\Delta_{95} = 2\sigma \left( \Delta z_{norm} \right)$. In our case, however, photo-z are empirical, i.e. not based on any specific fitting model and it is preferable to use the standard deviation value $\sigma \left( \Delta z_{norm} \right) $  derived from the photometric cross matched samples, although it could overestimate the theoretical Gaussian $\sigma$, due to the residual spectroscopic uncertainty as well as to the method training error. Therefore, we consider as catastrophic outliers the objects with $\left| \Delta z_{norm} \right| > 2 \sigma \left( \Delta z_{norm} \right)$. It is also important to notice that for empirical methods it is useful to analyze the correlation between the $NMAD\left( \Delta z_{norm} \right) = 1.48 \times median \left( \left| \Delta z_{norm} \right| \right)$ and the standard deviation $\sigma_{clean}(\Delta z_{norm})$ calculated on the data sample for which $\left| \Delta z_{norm} \right| \leq 2 \sigma \left( \Delta z_{norm} \right)$. In fact, the quantity $NMAD$ would be comparable to the value of the $\sigma_{clean}$.

\begin{figure}
\centering
  \includegraphics[width=10.cm]{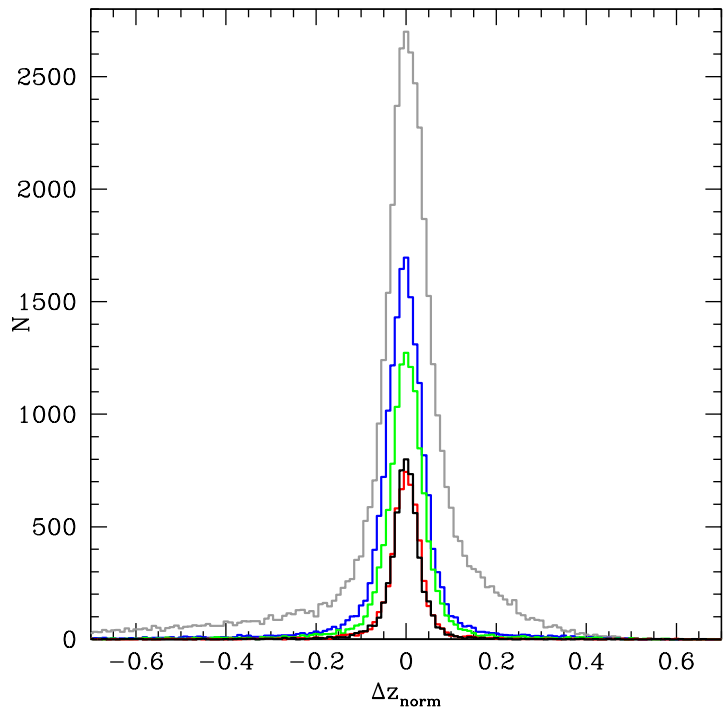}
\caption{$\Delta z_{norm}$ distributions for all five cross-matched test data sets. Lines are referred to, respectively, SDSS (gray), SDSS+GALEX (blue), SDSS+UKIDSS (green), SDSS+GALEX+UKIDSS (red) and SDSS+GALEX+UKIDSS+WISE (black).\label{qso:fig:histogram}}
\end{figure}

As it is shown in Tab.~\ref{qso:tab:catastrop}, in our data the $\sigma_{clean}(\Delta z_{norm})$ is always slightly larger than the corresponding $NMAD(\Delta z_{norm})$, which is exactly what is expected due to the overestimate induced by the above considerations (see also Fig.~\ref{qso:fig:histogram}).

Finally, we would like to stress that the difficulties  encountered by us and by other teams in comparing different methods, especially in light of the crucial role that photo-z play in the scientific exploitation of present and future large surveys (cf. \citealt{des2005}, \citealt{chambers2011}, \citealt{refregier2010}), confirm that it would be desirable to re-propose an upgraded version of the extremely useful PHAT contest (\citealt{hildebrandt2010}, \citealt{cavuoti2012b}), which allowed a direct, effective and non ambiguous comparison of different methods applied on the same datasets and through the same set of statistical indicators.
This new contest should be applied to a much larger dataset, with a more practical selection of photometric bands, and should take into account also other parameters such as scalability and robustness of the algorithms, as well as the degeneracy characterization.

\acknowledgments
\subsection*{Acknowledgments}
\noindent The authors would like to thank the anonymous referee for the comments and suggestions which
 helped us to improve the paper.
\noindent The authors wish to thank the whole DAMEWARE team\footnote{\emph{http://dame.dsf.unina.it/project\_members.html}},
for the many useful discussions.\\
\noindent The authors also wish to thank the financial support of Project F.A.R.O., $3^{rd}$ call by the University Federico II of Naples, and of the PRIN-MIUR 2011 for Euclid Mission.\\
\noindent One of us (GL) wishes to thank Prof G.S. Djorgovski and the whole Department of Astronomy at the California Institute of Technology in Pasadena, for hospitality.\\
\noindent AM and MB wish to thank the financial support of PRIN-INAF 2010, \textit{Architecture and Tomography of Galaxy Clusters}.\\
\noindent R. D'A. acknowledges the financial support of the US Virtual Astronomical Observatory, which is sponsored by the National Science Foundation and the National Aeronautics and Space Administration.\\

\clearpage

\begin{deluxetable}{ccll}
\tabletypesize{\scriptsize}
\tablewidth{0pt}
\tablecaption{Summary of the data extracted from the databases of the four surveys and merged to form our final catalogue. Even though most names of the parameters are self explanatory, we wish to remind that the various \textit{psfMag} are magnitudes derived by integrating fluxes over the best fitting point spread function. The aperture sizes refer to the radii.\label{qso:tab:features}}
\tablehead{\colhead{Survey} & \colhead{Bands} & \colhead{Name of feature} & \colhead{Synthetic description}}
\startdata
GALEX & nuv, fuv & mag, mag\_iso & Near and Far UV total and isophotal mags \\
& &mag\_Aper\_1 mag\_Aper\_2 mag\_Aper\_3& phot. through 3, 4.5 and 7.5 arcsec apertures\\
& & mag\_auto and  kron\_radius & magnitudes and Kron radius in units of A or B\\
\tableline
SDSS       & u, g, r, i, z  & psfMag& PSF fitting magnitude in the u g, r, i, z bands.\\
\tableline
UKIDSS   & Y, J, H, K  & PsfMag                                & PSF fitting magnitude in $Y, J, H, K$ bands\\
              & & AperMag3, AperMag4, AperMag6                 & aperture photometry through 2,   2.8 \& 5.7$^{\prime\prime}$\\
                    &&&  circular aperture in each band\\
             & & HallMag, PetroMag                                               & Calibrated magnitude within circular \\
             &&&aperture r\_hall and Petrosian magnitude\\
             & & & in $Y, J, H, K$ bands\\
\tableline
WISE           & W1, W2, W3, W4 & W1mpro, W2mpro, W3mpro, W4mpro
                                                                    & W1: 3.4 $\mu m$ and 6.1$^{\prime\prime}$ angular resolution;\\
                                                                &&& W2: 4.6 $\mu m$ and 6.4$^{\prime\prime}$ angular resolution; \\
                                                                &&&W3: 12 $\mu m$ and 6.5$^{\prime\prime}$ angular resolution;\\
                                                                &&&W4: 22 $\mu m$ and 12$^{\prime\prime}$ angular resolution.\\
                                                                &&& Magnitudes measured with profile-fitting photometry \\	
                                                                &&& at the 95\% level. Brightness upper limit if the flux \\
                                                                &&&measurement has SNR$<2$\\
 \tableline
 SDSS & - & $z_{spec}$ & Spectroscopic redshift\\
 \tableline
\enddata
\end{deluxetable}

\clearpage

\begin{deluxetable}{cccccc}
\tabletypesize{\scriptsize}
\tablecaption{Experiments for the feature selection phase.  Col.s 1-4: surveys used for the experiment, where superscript index indicates the used magnitudes: $^{1}$ \textit{mag}; $^{2}$ \textit{mag$\_$iso}; $^{3}$ \textit{magnitudes through 3, 4.5 and 7.5 arcsec apertures}; $^{4}$ \textit{mag$\_$auto}; $^{5}$ \textit{kron\_radius}; $^{6}$ \textit{HallMag}; $^{7}$ \textit{PetroMag}. A cross in a column means that the survey corresponding to that column was used for the experiment.\label{qso:tab:experiments}}
\tablewidth{0pt}
\tablehead{
\colhead{GALEX} & \colhead{SDSS} & \colhead{UKIDSS} & \colhead{WISE} & \colhead{$bias\left(\Delta z \right)$} & \colhead{$\sigma \left(\Delta z \right)$}}
\startdata
\cutinhead{Service Experiments}
 X				&	 X			&	 X			&	 X			&	    0.0033 	&	 0.174 		 \\
 X$^{1,2}$ 		&	 X			&	 X$^{6}$ 	&	 X			&	  -0.0001 	&	 0.152 		 \\
 X$^{3}$ 		&	 X			&	 X$^{6}$ 	&	 X		   	&	 -0.0016 	&	 0.165 		 \\
 X$^{1}$ 		&	 X			&	 X$^{6}$ 	&	 X			&	  0.0054 	&	 0.151 		 \\
 X$^{2}$ 		&	 X			&	 X$^{6}$ 	&	 X         	&	 -0.0026 	&	 0.151 		 \\
 X$^{4,5}$ 		&	 X			&	 X$^{6}$ 	&	 X			&	 -0.0008 	&	 0.152 		 \\
 X$^{1,2,3}$ 	&	 X			&	 X$^{6}$ 	&	 X			&	  0.0041 	&	 0.163 		 \\
 X$^{2,3}$		&	 X			&	 X$^{6}$ 	&	 X			&	  -0.0033 	&	 0.155 		 \\
 				&	 			&	 X$^{6,7}$ 	&	        	&	 -0.0091 	&	 0.299 		 \\
 				&	 			&	 X$^{7}$  	&	           	&	 0.0111 	&	 0.465 		 \\	
 				&	 			&	 X$^{6}$  	&	           	&	 -0.0081 	&	 0.294 		\\

\tableline
\enddata

\end{deluxetable}

\clearpage

\addtocounter{table}{-1}

\begin{deluxetable}{cccccc}
\tabletypesize{\scriptsize}
\tablecaption{continue}
\tablewidth{0pt}
\tablehead{
\colhead{GALEX} & \colhead{SDSS} & \colhead{UKIDSS} & \colhead{WISE} & \colhead{$bias\left(\Delta z \right)$} & \colhead{$\sigma \left(\Delta z \right)$}}
\startdata
\cutinhead{Four Survey Experiment}
 X$^{2}$ 		&	 X			&	 X$^{6}$ 	&	 X         	&	 -0.0026 	&	 0.151 		 \\
\cutinhead{Three Survey Experiment}
 X$^{2}$ 		&	 X			&	 X$^{6}$ 	&	           	&	 -0.0046 	&	 0.152 		 \\
 X$^{2}$ 		&	 X			&	 			&	 X         	&	 0.0025 	&	 0.162 		 \\
  				&	 X			&	 X$^{6}$ 	&	 X         	&	 -0.0032 	&	 0.179 		 \\
 X$^{2}$ 		&	 			&	 X$^{6}$ 	&	 X         	&	 0.0110 	&	 0.203 		 \\
\cutinhead{Two Survey Experiment}
		 		&	 			&	 X$^{6}$ 	&	 X         	&	 0.0045 	&	 0.236 		 \\
 X$^{2}$ 		&	 			&	  			&	 X         	&	 0.0175 	&	 0.288 		 \\
  				&	 X			&	 X$^{6}$ 	&	         	&	 -0.0027 	&	 0.210 		 \\
  				&	 X			&	 			&	 X        	&	 -0.0039 	&	 0.197 		  \\
 X$^{2}$ 		&	 X			&	  			&            	&	 -0.0055 	&	 0.240 		 \\
 X$^{2}$ 		&	  			&	 X$^{6}$ 	&	         	&	 0.0133 	&	 0.238 		\\
 \cutinhead{One Survey Experiment}
 				&	 			&	 			&	 X         	&	 0.0165 	&	 0.297 		\\
 				&	 X  		&		 		&	           	&	 -0.0162 	&	 0.338 		 \\
 X$^{1,2}$ 		&	 			&	  			&			  	&	 0.0550 	&	 0.419 		 \\	
 				&	 			&	 X$^{6}$  	&	           	&	 -0.0081 	&	 0.294 		\\
\tableline
\enddata
\tablenotetext{1} {\textit{mag}}
\tablenotetext{2} {\textit{mag$\_$iso}}
\tablenotetext{3} {\textit{mag\_Aper 1, 2 and 3}}
\tablenotetext{4} {\textit{mag$\_$auto}}
\tablenotetext{5} {\textit{kron\_radius}}
\tablenotetext{6} {\textit{HallMag}}
\tablenotetext{7} {\textit{PetroMag}}
\end{deluxetable}

\begin{deluxetable}{ccccc}
\tabletypesize{\scriptsize}
\centering
\tablecaption{Comparison among the performances of the different references. MLPQNA is related to our experiments, based on a four-layers network, trained on the mixed (colors + reference magnitudes) datasets. In some cases the comparison references are not reported, due to the missing statistics. Column 1: reference; Column 2-5, respectively: bias, standard deviation, MAD, RMS, calculated on $\Delta z = \left( z_{spec}-z_{phot}\right)$ related to the test sets. For the definition of the parameters and for discussion see text.\label{qso:tab:comparison1}}
\tablewidth{0pt}
\tablehead{
\colhead{Exp} & \colhead{$BIAS(\Delta z)$} & \colhead{$\sigma(\Delta z)$} & \colhead{$MAD(\Delta z)$} & \colhead{$RMS(\Delta z)$}}
\startdata
\cutinhead{SDSS}
MLPQNA          & 0.007  & 0.25 & 0.102 & 0.26\\
Bovy et al.     & -      & 0.46 & -     & -\\
Laurino et al.  & 0.210  & 0.28 & 0.110 & 0.35\\
Ball et al.     & -      & 0.35 & -     & -\\
Richards et al. & -      & 0.52 & -     & -\\
\cutinhead{SDSS + GALEX}
MLPQNA          & 0.003  & 0.21 & 0.060 & 0.22\\
Bovy et al.     & -      & 0.26 & -     & -\\
Laurino et al.  & 0.13   & 0.21 & 0.061 & 0.25\\
Ball et al.     & -      & 0.23 & -     & -\\
Richards et al. & -      & 0.37 & -     & -\\
\cutinhead{SDSS + UKIDSS}
MLPQNA          & 0.001  & 0.25 & 0.066 & 0.26\\
Bovy et al.     & -      & 0.28 & -     & -\\
\cutinhead{SDSS + GALEX + UKIDSS}
MLPQNA          & 0.0009 & 0.18 & 0.043 & 0.19\\
Bovy et al.     & -      & 0.21 & -     & -\\
\cutinhead{SDSS + GALEX + UKIDSS + WISE}
MLPQNA          & 0.002  & 0.15 & 0.040 & 0.15\\
\tableline
\enddata
\end{deluxetable}

\clearpage

\begin{deluxetable}{cccccc}
\tabletypesize{\scriptsize}
\centering
\tablecaption{Comparison among the performances of the different references. MLPQNA is related to our experiments, based on a four-layers network, trained on the mixed (colors + reference magnitudes) datasets. In some cases the comparison references are not reported, due to the missing statistics. Column 1: reference; columns 2-6, respectively: bias, standard deviation, MAD, RMS and NMAD calculated on $\Delta z_{norm} = \left( z_{spec}-z_{phot}\right)/\left(1 + z_{spec}\right)$ related to the test sets. For the definition of the parameters and for discussion see text.\label{qso:tab:comparison2}}
\tablewidth{0pt}
\tablehead{
\colhead{Exp} & \colhead{$BIAS(\Delta z_{norm})$} & \colhead{$\sigma(\Delta z_{norm})$} & \colhead{$MAD(\Delta z_{norm})$} & \colhead{$RMS(\Delta z_{norm})$} & \colhead{$NMAD(\Delta z_{norm})$}}
\startdata
\cutinhead{SDSS}
MLPQNA          & 0.032  & 0.15 & 0.039 & 0.17 &0.058\\
Laurino et al.  & 0.095  & 0.16 & 0.041 & 0.19 &-\\
Ball et al.     & 0.095  & 0.18 & -     & -    &-\\
Richards et al. & 0.115  & 0.28 & -     & -    &-\\
\cutinhead{SDSS + GALEX}
MLPQNA          & 0.012  & 0.11 & 0.029 & 0.11 &0.043\\
Laurino et al.  & 0.058  & 0.29 & 0.029 & 0.11 &-\\
Ball et al.     & 0.06   & 0.12 & -     & -    &-\\
Richards et al. & 0.071  & 0.18 & -     & -    &-\\
\cutinhead{SDSS + UKIDSS}
MLPQNA          & 0.008  & 0.11 & 0.027 & 0.11 &0.040\\
\cutinhead{SDSS + GALEX + UKIDSS}
MLPQNA          & 0.005  & 0.087 & 0.022 & 0.088 &0.032\\
\cutinhead{SDSS + GALEX + UKIDSS + WISE}
MLPQNA          & 0.004  & 0.069 & 0.020 & 0.069 & 0.029\\
\tableline
\enddata
\end{deluxetable}

\clearpage

\begin{deluxetable}{ccccc}
\tabletypesize{\scriptsize}
\centering
\tablecaption{Comparison in terms of outliers percentages among the different references. In some cases the comparison references are not reported, due to the missing statistics. Column 1: reference; Column 2-3 are fractions of outliers at different $\sigma$ based on $\Delta z = \left( z_{spec}-z_{phot}\right)$; Column 4-5 are the fractions of outliers at different $\sigma$ based on $\Delta z_{norm} = \left( z_{spec}-z_{phot}\right)/\left(1 + z_{spec}\right)$. The column 4 reports our catastrophic outliers, defined as $|\Delta z_{norm}|>2\sigma(\Delta z_{norm})$.\label{qso:tab:compoutliers}}
\tablewidth{0pt}
\tablehead{
\colhead{Exp} & \colhead{Outliers ($|\Delta z|$)} & & \colhead{Outliers ($|\Delta z_{norm}|$)}}
\startdata
  	 & 	$>2\sigma(\Delta z)$ & $> 4\sigma(\Delta z)$ 	& $>2\sigma(\Delta z_{norm})$ & $> 4\sigma(\Delta z_{norm})$ \\
\cutinhead{SDSS}
MLPQNA 			&				 7.68 		&	 0.38 			&	 6.53 			& 			 1.24\\
Bovy et al. 	&							&	 0.51	\\
\cutinhead{SDSS + GALEX}
MLPQNA 			&				 4.88 		&	 1.61 			&	 4.57			&			1.37\\
Bovy et  al. 	&							&	 1.86	\\
\cutinhead{SDSS + UKIDSS}
MLPQNA 			&				4.00 		&	 1.73			&	 3.82			&			 1.38\\
Bovy et al. 	&							&	 1.92	\\
\cutinhead{SDSS + GALEX + UKIDSS}
MLPQNA 			&				2.86 		& 	 1.47 			&	 3.05			&			0.23\\
Bovy et al. 	&							&	 1.13 	\\
\cutinhead{SDSS + GALEX + UKIDSS + WISE}
MLPQNA 			&				 2.57		&	 0.87			&	 2.88			&			 0.91\\
\tableline
\enddata
\end{deluxetable}

\clearpage

\begin{deluxetable}{lccccc}
\tabletypesize{\scriptsize}
\tablecaption{Catastrophic outliers evaluation and comparison between the residual $\sigma_{clean}(\Delta z_{norm})$ and $NMAD(\Delta z_{norm})$. The reported number of objects, for each cross-matched catalog, is referred to the test sets only. Catastrophic outliers are defined as objects where $\left| \Delta z_{norm} \right| > 2 \sigma \left( \Delta z_{norm} \right)$. The standard deviation $\sigma_{clean}(\Delta z_{norm})$ is calculated after having removed the catastrophic outliers, i.e. on the data sample for which $\left| \Delta z_{norm} \right| \leq 2 \sigma \left( \Delta z_{norm} \right)$ \label{qso:tab:catastrop}}
\tablewidth{0pt}
\tablehead{
\colhead{Exp} & \colhead{n. obj.} & \colhead{$\sigma \left( \Delta z_{norm} \right) $} & \colhead{\% catas. outliers} & \colhead{$\sigma_{clean} \left( \Delta z_{norm} \right)$} &
\colhead{$NMAD\left( \Delta z_{norm} \right)$}}
\startdata
SDSS                   & 41431 & 0.15  & 6.53 & 0.062 & 0.058\\
SDSS + GALEX           & 17876 & 0.11  & 4.57 & 0.045 & 0.043\\
SDSS+UKIDSS            & 12438 & 0.11  & 3.82 & 0.041 & 0.040\\
SDSS+GALEX+UKIDSS      & 5836  & 0.087 & 3.05 & 0.040 & 0.032\\
SDSS+GALEX+UKIDSS+WISE & 5716  & 0.069 & 2.88 & 0.035 & 0.029\\
\enddata
\end{deluxetable}

\clearpage

\begin{figure}
\centering
  \includegraphics[width=12.cm]{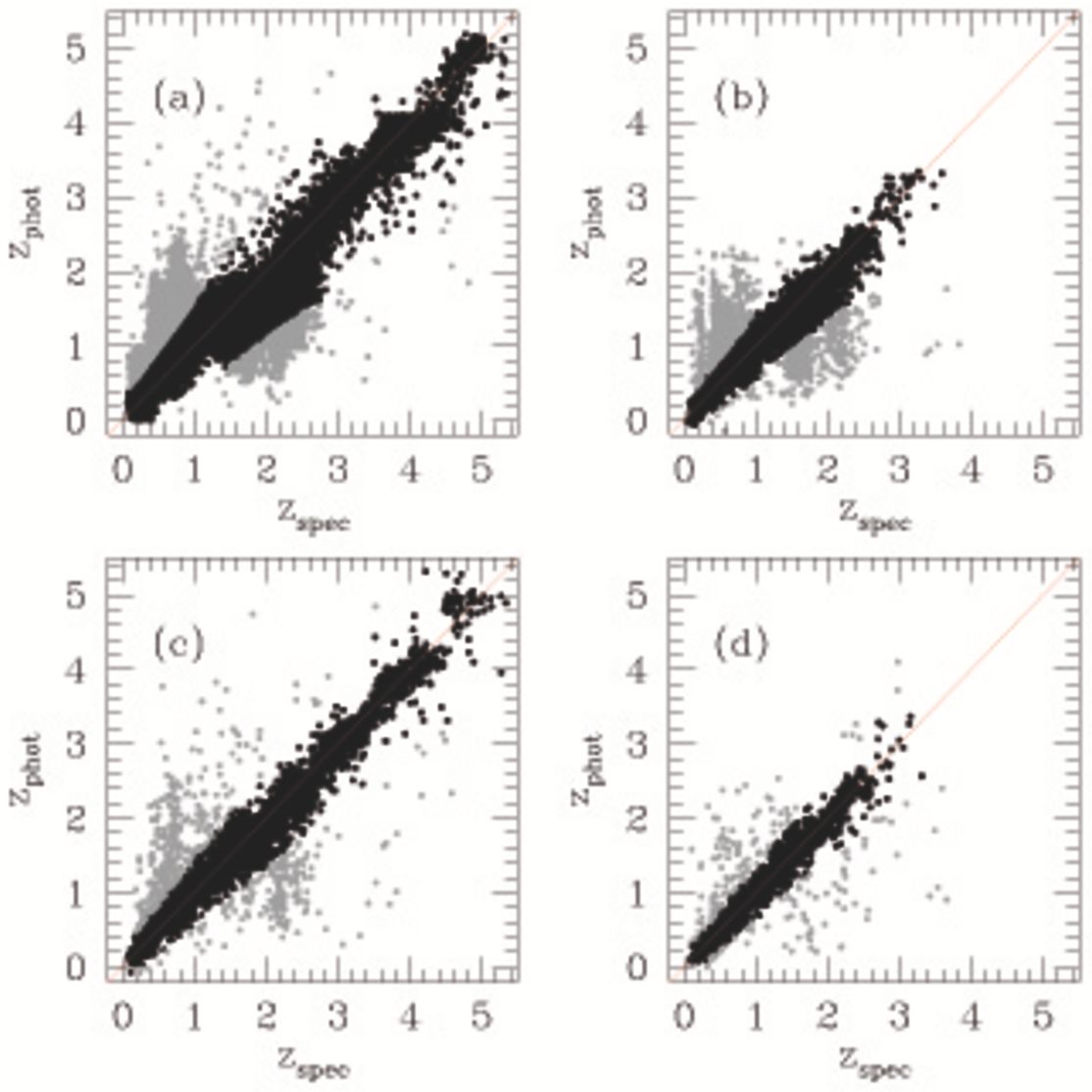}
  \includegraphics[bb= 1 220 320 560, clip,width=7.cm]{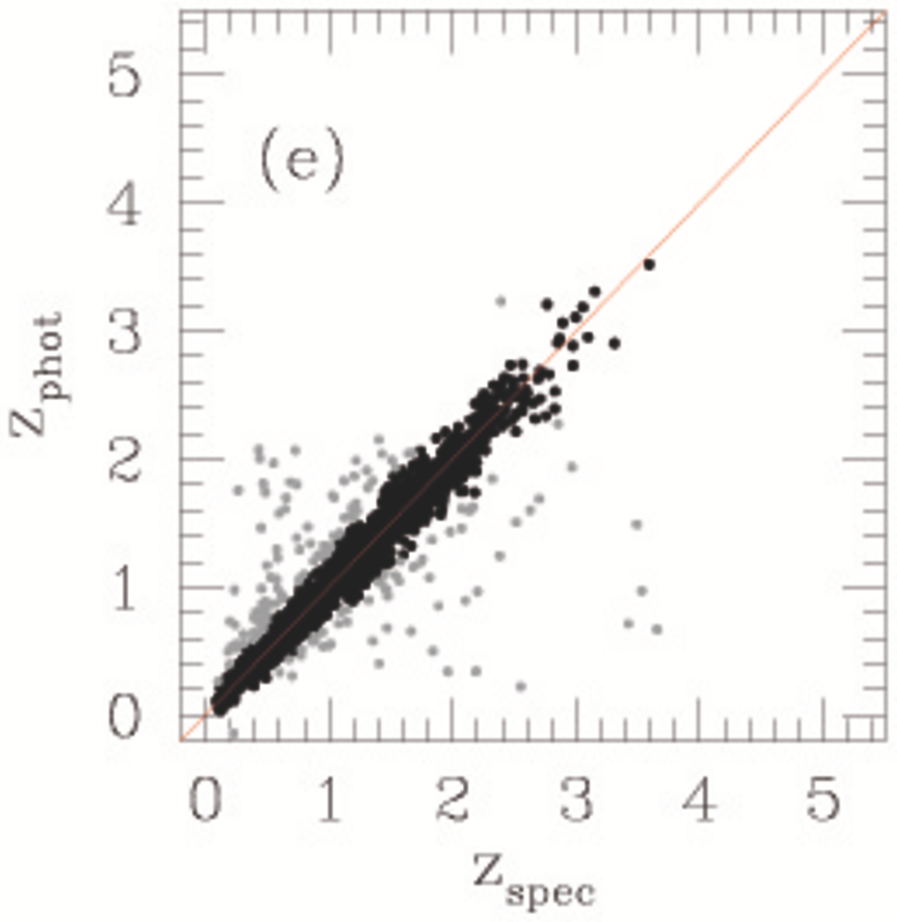}
\caption{Scatter plots ($z_{spec}$ vs $z_{phot}$); (a) SDSS, (b) SDSS+GALEX, (c) SDSS+UKIDSS, (d) SDSS+GALEX+UKIDSS and (e) SDSS+GALEX+UKIDSS+WISE. All diagrams refer to results on test sets. Gray points are catastrophic outliers (defined in Tab.~\ref{qso:tab:compoutliers}). Red line is the dot-to-dot straight line passing through photometric and spectroscopic redshift limits in the available Knowledge Base.\label{qso:fig:SCATTERtestset}}
\end{figure}






\clearpage

\appendix

\section{Appendix: The Quasi Newton learning rule}
\label{qso:appendixA}

Most Newton methods use the Hessian of the function to find the stationary point of a quadratic form. It needs to be stressed, however, that the Hessian of a function is not always available and in many cases it is far too complex to be computed in an analytical way.
More often it is easier to compute the function gradient which can be used to approximate the Hessian via $N$ consequent gradient calculations.
In order to better understand why QNA are so powerful, it is convenient to start from the classical and quite common  Gradient Descent Algorithm (GDA) used for Back Propagation \citep{bishop2006}. In GDA, the direction of each updating step for the MLP weights is derived from the error descent gradient, while the length of the step is determined from the learning rate. In case of particularly complex problems this method is inaccurate and ineffective and therefore may get stuck in local minima.
A more effective approach is to move towards the negative direction of the gradient (\textit{line search direction}) not by a fixed step, but by moving towards the minimum of the function along that direction.  This can be achieved by first deriving the descent gradient and then by analyzing it with the variation of the learning rate \citep{brescia2012d}.
Let us suppose that at step $t$, the current weight vector is $w^{(t)}$, and let us consider a search direction $d^{(t)} =  - \nabla E^{(t)}$.
If we select the parameter $\lambda$ in order to minimize $E(\lambda) = E({w^{(t)} + \lambda d^{(t)} })$, the new weight vector can be expressed as:

\begin{equation}
w^{({t + 1})}  = w^{(t)}  + \lambda d^{(t)}
\end{equation}\label{qso:eq1}

\noindent and the problem of \textit{line search} becomes a  1-dimensional minimization problem which can be solved in many different ways. Simple variants are: i) to move $E(\lambda)$ by varying $\lambda$ by small intervals, then evaluate the error function at each new position, and stop when the error begins to increase, or ii) to use the parabolic search for a minimum and compute the parabolic curve crossing pre-defined learning rate points. The minimum $d$ of the parabolic curve is a good approximation of the minimum of $E(\lambda)$ and it can be derived by means of the parabolic curve which crosses the fixed points with the lowest error values.

Another approach makes instead use of \textit{trust region} based strategies which minimize the error function, by iteratively growing or contracting the region of the function by adjusting a quadratic model function which best approximates the error function. In this sense this technique can be considered as a dual to line search, since it tries to find the best size of the region by fixing the step size (while the line search strategy always chooses the step direction before selecting the step size), \citep{celis1985}.
All these approaches, however, rely on the assumption that the optimal search direction is given at each step by the negative gradient: an assumption which not only is not always true, but can also lead to serious wrong convergence.
In fact, if the minimization is done along the negative gradient direction, the subsequent search direction (the new gradient) will be orthogonal to the previous one: in fact, note that when the line search founds the minimum, it is:
\begin{equation}
\frac{{\partial E}}{{\partial \lambda }}({w^{(t)} + \lambda d^{(t)} }) = 0
\label{qso:eq2}
\end{equation}
and hence,
\begin{equation}
g^{({t + 1})T} d^{(t)} = 0
\label{qso:eq3}
\end{equation}
where $g \equiv \nabla E$.
The iteration of the process therefore leads to oscillations of the error function which slow down the convergence process. The method implemented here relies on selecting other directions so that the gradient component, parallel to the previous search direction, would remain unchanged at each step. Suppose that you have already minimized with respect to the direction $d^{(t)}$ starting from the point $w^{(t)}$ and reaching the point $w^{(t+1)}$, where Eq. \ref{qso:eq3} becomes:
\begin{equation}
g({w^{({t + 1})}})^T d^{(t)} = 0
\label{qso:eq4}
\end{equation}
\noindent by choosing $d^{(t+1)}$ so to preserve the gradient component parallel to $d^{(t)}$ equal to zero, it is possible to build a sequence of directions $d$ in such a way that each direction is conjugated to the previous one on the dimension $|w|$ of the search space (this is known as conjugate gradients method; \cite{golub1999}).
In presence of a squared error function, the update weights algorithm is:
\begin{equation}
w^{({t + 1})} = w^{(t)} + \alpha ^{(t)} d^{(t)}
\label{qso:eq5}
\end{equation}
with:
\begin{equation}
\alpha ^{(t)} = - \frac{{d^{(t)T} g^{(t)} }}{{d^{(t)T} Hd^{(t)} }}
\label{qso:eq6}
\end{equation}

Furthermore, $d$ can be obtained for the first time via the negative gradient and in the subsequent iterations,  as a linear combination of the current gradient and of the previous search directions:
\begin{equation}
d^{({t + 1})} = - g^{({t + 1})} + \beta ^{(t)} d^{(t)}
\label{qso:eq7}
\end{equation}
with:
\begin{equation}
\beta ^{(t)} = \frac{{g^{({t + 1})T} Hd^{(t)} }}{{d^{(t)T} Hd^{(t)} }}
\label{qso:eq8}
\end{equation}

This algorithm finds the minimum of a square error function at most in $|w|$ steps but at the price of a high computational cost, since in order to determine the values of $\alpha$ and $\beta$, it makes use of that \textit{hessian matrix H} which, as we already mentioned, is very demanding in terms of computing time. A fact which puts serious constraints on the application of this family of methods to medium/large data sets.  Excellent approximations for the coefficients $\alpha$ and $\beta$ can, however, be obtained from analytical expressions that do not use the Hessian matrix explicitly.
For instance, $\beta$ can be calculated through any one of the following expressions (respectively \cite{hestenes1952, fletcher1964, polak1969}):

\begin{equation}
Hestenes-Sitefel: \beta ^{(t)} = \frac{{g^{({t + 1})T}( {g^{( {t + 1})} - g^{(t)}})}}{{d^{(t)T}({g^{({t + 1})} - g^{(t)}})}}
\label{qso:eq9}
\end{equation}
\begin{equation}
Fletcher-Reeves: \beta ^{(t)} = \frac{{g^{({t + 1})T} g^{({t + 1})} }}{{g^{(t)T} g^{(t)} }}
\label{qso:eq10}
\end{equation}
\begin{equation}
Polak-Ribiere: \beta ^{(t)} = \frac{{g^{({t + 1})T} ( {g^{({t + 1})} - g^{(t)} })}}{{g^{(t)T}g^{(t)}}}
\label{qso:eq11}
\end{equation}

\noindent These expressions are all equivalent if the error function is square-typed, otherwise they assume different values. Typically the Polak-Ribiere equation obtains better results because, if the algorithm is slow and subsequent gradients are quite alike between them, its equation produces values of $\beta$ such that the search direction tends to assume the negative gradient direction \citep{vetterling1992}.

Concerning the parameter $\alpha$, its value can be obtained by using the line search method directly. The method of conjugate gradients reduces the number of steps to minimize the error up to a maximum of $|w|$ because there could be almost $|w|$ conjugate directions in a $|w|$-dimensional space. In practice however, the algorithm is slower because, during the learning process, the property \textit{conjugate} of the search directions tends to deteriorate.
It is useful, to avoid the deterioration, to restart the algorithm after $|w|$ steps, by resetting the search direction with the negative gradient direction.

By using a local square approximation of the error function, we can obtain an expression for the minimum position. The gradient in every point $w$ is in fact given by:
\begin{equation}
\nabla E = H \times ( {w - w^* })
\label{qso:eq12}
\end{equation}
where $w^*$ corresponds to the minimum of the error function, which satisfies the condition:
\begin{equation}
w^* = w - H^{ - 1} \times \nabla E
\label{qso:eq13}
\end{equation}

The vector $- H^{ - 1}  \times \nabla E$ is known as Newton direction and it is the base for a variety of optimization strategies, such as for instance the QNA, which instead of calculating the $H$ matrix and then its inverse, uses a series of intermediate steps of lower computational cost to generate a sequence of matrices which are more and more accurate approximations of $H^{ - 1}$. From the Newton formula (Eq. \ref{qso:eq13}) we note that the weight vectors on steps $t$ and $t+1$ are correlated to the correspondent gradients by the formula:
\begin{equation}
w^{({t + 1})} - w^{(t)} = - H^{({ - 1})}( {g^{({t + 1})} - g^{(t)}})
\label{qso:eq14}
\end{equation}
which is known as \textit{Quasi Newton Condition}. The approximation $G$ is therefore built in order to satisfy this condition. The formula for $G$ is:
\begin{equation}
G^{({t + 1})} = G^{(t)} + \frac{{pp^T }}{{p^T \nu }} - \frac{{({G^{(t)} \nu })\nu ^T G^{(t)}}}{{\nu ^T G^{(t)} \nu }} + ( {\nu ^T G^{(t)} \nu })uu^T
\label{qso:eq15}
\end{equation}
where the vectors are:
\begin{equation}
p = w^{({t + 1})} - w^{(t)}; \nu = g^{( {t + 1})} - g^{(t)}; u = \frac{p}{{p^T \nu }} - \frac{{G^{(t)} \nu }}{{\nu ^T G^{(t)} \nu }}
\label{qso:eq16}
\end{equation}

Using the identity matrix to initialize the procedure is equivalent to consider, step by step, the direction of the negative gradient while, at each next step, the direction $-Gg$ is for sure a descent direction. The above expression could carry the search out of the interval of validity for the squared approximation. The solution is hence to use the \textit{line search} to found the minimum of function along the search direction.
By using such system, the weight updating expression (Eq. \ref{qso:eq5}) can be formulated as follows:
\begin{equation}
w^{({t + 1})} = w^{(t)} + \alpha ^{(t)} G^{(T)} g^{(t)}
\label{qso:eq17}
\end{equation}
where $\alpha$ is obtained by the \textit{line search}.

One of the main advantages of QNA, compared with conjugate gradients, is that the \textit{line search} does not require the calculation of $\alpha$ with a high precision, because it is not a critical parameter. Unfortunately, however, again, it requires a large amount of memory to calculate the matrix $G$ ($|w| \times |w|$), for large $|w|$.
One way to reduce the required memory is to replace at each step the matrix $G$ with a unitary matrix. With such replacement and after multiplying by $g$ (the current gradient), we obtain:
\begin{equation}
d^{({t + 1})} = - g^{(t)} + Ap + B\nu
\label{qso:eq18}
\end{equation}

Note that if the line search returns exact values, then the above equation produces mutually conjugate directions. $A$ and $B$ are scalar values defined as:
\begin{equation}
\begin{array}{l}
 A =  - ({1 + \frac{{\nu ^T \nu }}{{p^T \nu }}})\frac{{p^T g^{({t + 1})} }}{{p^T \nu }} + \frac{{\nu ^T g^{({t + 1})} }}{{p^T \nu }} \\
 \\
 B = \frac{{p^T g^{({t + 1})} }}{{p^T \nu }} \\
 \end{array}
\label{qso:eq19}
\end{equation}

\end{document}